\documentclass[journal,10pt]{IEEEtran}

\pdfoutput=1

\usepackage{relsize}
\usepackage{amssymb}
\usepackage{amsmath}
\usepackage{blindtext, graphicx}
\usepackage{tabularx}
\usepackage{xcolor}
\usepackage{soul}
\usepackage{cite}
\usepackage{bigints}
\usepackage{algorithm}
\usepackage{algorithmic}
\usepackage{float}
\usepackage{stfloats}
\usepackage{cuted}
\usepackage{etoolbox}
\AfterEndEnvironment{strip}{\leavevmode}

\begin{document}
%

\title{Statistical Modelling of Dynamic Interference Threshold and Its Effect on Network Capacity}

\author{
Amit~Kachroo,~\IEEEmembership{Student~Member,~IEEE,}  
Sabit~Ekin,~\IEEEmembership{Member,~IEEE,} and
Ali~Imran,~\IEEEmembership{Senior~Member,~IEEE} 

\thanks{Copyright (c) 2015 IEEE. Personal use of this material is permitted. However, permission to use this material for any other purposes must be obtained from the IEEE by sending a request to pubs-permissions@ieee.org. }

\thanks{ A.~Kachroo, and S.~Ekin are with the School of Electrical and Computer Engineering, Oklahoma State University, OK, USA (e-mail: (amit.kachroo, sabit.ekin)@okstate.edu).}
\thanks{A.~Imran is with the Telecommunications Engineering, University of Oklahoma, OK, USA (e-mail:~ali.imran@ou.edu).}
}

\markboth{Accepted for publication in IEEE Transactions on Vehicular Technology,~Vol.~XX, No.~XX, XXX~2020}
{}%

\maketitle


\begin{abstract}

In this paper, we present the case of utilizing interference temperature (IT)  as a dynamic quantity rather than as a fixed quantity in an  orthogonal frequency division multiple access (OFDMA) based spectrum sharing systems. The fundamental idea here is to reflect the changing capacity demand of primary user (PU) over time in setting the interference power threshold for secondary user (SU).  This type of  dynamic IT will allow  the SU to opportunistically have higher transmit power  during relaxed IT period, thereby resulting in higher network capacity.  The cognitive radio network (CRN) considered in this paper has an underlay network configuration in which the available spectrum of the PU is accessed concurrently by SU provided  that the interference power at the PU receiver from SU is under a certain power threshold. This power threshold is set to maintain and guarantee a certain level of quality of service (QoS) for PU network.  Theoretical expressions for outage probability and mean capacity for SU network are derived, and validated with simulation results,  and it is observed that utilizing dynamic IT results in high network performance  gains as compared to utilizing a fixed IT in cognitive radio system.
\end{abstract}


%
\IEEEpeerreviewmaketitle

\section{Introduction}
The explosive growth of mobile devices over the last decade has  created a lot of stress on the available frequency spectrum  for public use. With more and more devices coming into the picture, the spectrum is getting more crowded than before. In the past decade or so, measurement studies on the actual spectrum have revealed that a large portion of the licensed spectrum is less utilized than the unlicensed one. These studies also highlight that this inefficient and inflexible spectrum allocation leads to the spectrum scarcity{\cite{datla2009spectrum,islam2008spectrum,federal2002spectrum}}.
To overcome this challenge of overcrowded spectrum, cognitive radio theory was introduced, where  radios in unlicensed spectrum would exploit the available licensed spectrum opportunistically, and thereby yield  high network efficiency\cite{mitola1998cognitive,mitola1999cognitive,haykin2005cognitive,erpek2007spectrum,cao_new}.

In cognitive radio network (CRN), the users of the radio spectrum are divided into two categories: licensed and unlicensed users. Depending on the network configuration, the unlicensed users  or the secondary users (SUs) are allowed to access the spectrum of the licensed users or the primary users (PUs) either when it is not in use by PU or concurrently with PU transmissions. The concurrent transmission is allowed, if and only if the SUs can maintain a certain interference power threshold such that it doesn't affect the  quality of service (QoS) for PU transmission. The spectrum sharing network with concurrent access of the available spectrum is known as underlay cognitive network, while the network that allows spectrum access only during idle time  is known as overlay cognitive network \cite{haykin2005cognitive,sithamparanathan2012cognitive}. Moreover, the required interference power threshold to maintain the certain QoS at PU-Rx is defined as interference temperature (IT)\cite{kolodzy2006interference,ekin2012random,aissacapacity,kachroo2018impact}.  In this work, we consider the network to be in  underlay configuration. 

In the underlay network, the SUs  adapt  their  transmit  power  to maintain the required IT constraint.  To maintain IT, SU will either adapt its peak or average transmit power \cite{kachroo2018impact,PrimaryImpact,derivation,PeakVsAverage}. In  \cite{kangPeakAvg}, Kang \textit{et al}. have studied and derived the optimal power strategies for SU to maximize  outage and ergodic capacity under both (peak and average power) constraints. Similarly, in \cite{peakOnly}, Srinivasa \textit{et al.} has considered peak and average power adaptation  to maximize the SU signal to noise ratio (SNR) and capacity. However, peak power adaptation protects and guarantees  instantaneous interference prevention at  PU, and in many cases, the PU QoS would be limited by the instantaneous signal to interference plus noise ratio (SINR) at the receiver.  Therefore, in this work, peak power adaptation was considered, however, it is important to note that the insights from peak power adaptation will  still be valid even if average power adaptation was considered. This type of power adaptation scheme requires the knowledge of  channel state information (CSI) at SU-Tx, so that the SU can adapt their transmit power accordingly. Recent research studies have shown that this can be  achieved by utilizing feedback channels with  acknowledgment/non-acknowledgement (ACK/ NACK) packet information  or by detecting the transition of modulation and coding schemes (MCS)\cite{ack3,ack1,ack2,mcs1,mcs2}. In our work, this job is accomplished by a  central entity known as CBS (central base station), which periodically senses the CSI information of PUs and SUs in a CRN. This type of  CRN with CBS is also known as a centralized CRN system\cite{ack3,mcs1,mcs2}. Apart from sensing CSI, CBS also senses the primary network  activity,  and  controls  the SUs via dedicated sensing and control channels.  However, in reality, the CSI knowledge is not perfect\cite{li_new} but  since the main aim of this paper is to statistically highlight the advantages of dynamic IT over fixed IT, we have safely assumed perfect CSI knowledge at SU-Tx. The system model in that regard will be discussed in more detail in Section \ref{sec2}.

Traditionally, the interference power threshold or IT for SUs is kept constant, however,  some studies, such as \cite{chen2012performance} has thoroughly analyzed the concept of interference probability in a relay assisted CRN, assuming imperfect CSI. In \cite{chen2012performance}, the authors have quantified the performance of spectrum sharing cognitive relay networks in the presence of imperfect CSI with a metric termed as interference probability. This interference probability is found to be always equal to 0.75: the probability that the actual IT is higher than the estimated IT  when the CSI is imperfect. Further, the well-known main requirement of CRN is to maintain QoS of PUs (legacy users), while aiming at increasing spectral efficiency of the whole system by allowing SUs to access the spectrum. However, fixed IT constraint could be considered a strict requirement for satisfying this QoS of PUs.

The primary motivation behind this work is to relax IT constraint dynamically while considering capacity demand requirement of PUs, hence improving spectral efficiency (i.e., increase capacity of SUs) by allowing SUs to opportunistically transmit at higher power levels. As per our knowledge, this is the first paper to statistically model  the dynamic IT considering the dynamic PU traffic demand. Utilizing these dynamic IT settings, the IT for SUs can be relaxed during  less traffic time or ideal time in the PU network, which will  allow  the SUs to increase their transmit power and thereby further improve the overall network performance of CRN.  This dynamic setting and modelling  is discussed in more detail in Section \ref{sec3}. In Sections \ref{sec4} and \ref{sec5}, derivations for different performance expressions in general and in  high power region is discussed in detail. The simulation results and discussion is presented in Section \ref{sec6}, and finally, the conclusion and future work is mentioned in Section VII. In a nutshell, this statistical modelling of dynamic IT and its effect on network performance is the main contribution of this work. In general, the contributions of this paper can be summarized as follows:
\begin{itemize}
    \item Theoretical probability density function (PDF) and cumulative distribution function (CDF) expressions for  SINR from variable Poisson distributed capacity demand  of a PU is found and validated with simulation results. 
    \item Theoretical PDF and CDF of dynamic interference power threshold  are derived and checked   with simulations.
     \item Theoretical derivations of outage probability and mean capacity of SU  in general operation region, and  in high power region are found and validated with simulations.
    \item The performance of CRN with dynamic IT and conservative fixed IT is compared and discussed.
\end{itemize}

\section{System Model}\label{sec2}

\begin{figure}[t]
    \centering
    \includegraphics[width=1\columnwidth]{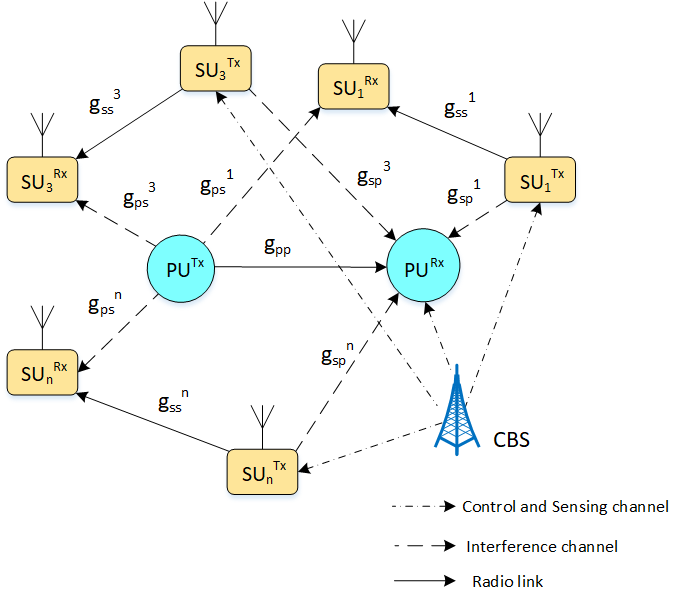}
    \caption{System model with $N$-SUs and a PU link.}
    \label{sysm}
\end{figure}

In this section, the system under consideration is discussed in detail. The system model is shown in Fig. \ref{sysm}, which consists of a PU network and $N$-SUs in an   underlay network configuration in which the available primary user spectrum is shared with $N$-SUs. Furthermore, it is assumed that the orthogonal frequency division multiple access (OFDMA) method is employed in the CRN that allows every PU to access orthogonal spectrum bands from the available bandwidth, and for each allocated PU orthogonal frequency  band, the SUs will operate in separate sub-bands of it. This will result in no  interference among SUs, but will cumulatively cause interference on the PU, whose spectrum is shared by these SUs. The various channel gains assuming point-to-point flat Rayleigh fading channels\footnote{
Rayleigh fading model is commonly used channel model for such theoretical studies. Other small-scale channel fading models such as Nakagami and Rician could be definitely considered, however, the insights and observations obtained from this study would remain the same.} are given as, 
$g_{sp}^1=||h_{sp}^1||^2$, $g_{ps}^1=||h_{ps}^1||^2$, $g_{ss}^1=||h_{ss}^1||^2$, and  $g_{pp}^1=||h_{pp}^1||^2$, where $g$ represents the channel power gain, $h$ represents the channel transfer function or channel response (Rayleigh), and subscript $p$ represents PU while subscript  $s$ represents SU.  Also, the superscripts  $1,\dots,n$ present the SU or PU index number, for example $g_{ss}^1$ is the channel between  the $SU^{Tx}_1$ and $SU^{Rx}_1$. Moreover, we  denote the exponentially distributed PDFs of these random variables as $f_{g_{sp}}(x)$, $f_{g_{ps}}(x)$, $f_{g_{ss}}(x)$ and $f_{g_{pp}}(x)$. These PDFs are governed by corresponding rate parameters,  which depend on the mean of the exponential distribution as $E(g_{sp})=1/ \lambda_{sp}$, $E(g_{ps})=1/ \lambda_{ps}$, $E(g_{ss})=1/ \lambda_{ss}$ and $E(g_{pp})=1/ \lambda_{pp}$. 

It is worth to note that the mean values of channel power parameters in small-scale fading models incorporate the effect of large-scale fading such as path-loss and shadowing under the assumption that there are immobile users, i.e., path-loss and shadowing will be constant\footnote{Note that the users are assumed to be immobile.}\cite{rayL,goldsmith2005wireless}. Consequently, a low mean  parameter would imply a larger distance between a PU-Rx and a SU-Tx than a high mean value; the mean value here refers to the received signal mean power over a distance. As an example, from Fig. \ref{sysm}, since  SU-1 is  nearer to PU-Rx than SU-3, it will have a high mean value ($E({g_{sp}}^1)$) than the SU-3. Therefore, selection of these rate parameters will take care of the distance dependency in itself. 

Apart from that, we also assume the channels to be flat in our model. Since  our main motivation was to show the network performance gain of utilizing dynamic IT over fixed IT, traffic scheduling and access control are not considered  in this work.  Moreover, we assume a CRN  of single PU with single SU for our analysis purposes, which  can be further  extrapolated to multiple SUs with a single PU case, or multiple PU case with different number of SUs\footnote{In case of multi-user scenario, the functionality of  CBS becomes critical as scheduling, channel allocations and users monitoring need to be performed by the CBS. For example, in this OFDMA-based system, CBS can allow only one SU to concurrently use the carrier-band with the PU. One can take insights from this study and carefully incorporate the operation of CBS to extend it to multi-user scenario.}. As mentioned in the earlier section, the centralized CRN \cite{ack3,mcs1,mcs2}  has  a CBS that controls the CRN operation by sensing the PU activity periodically via sensing channels, and  sets the dynamic IT  for SUs via control channels. It is also important to highlight here that the CSI knowledge can not be obtained perfectly in a practical wireless network, but with this centralized CRN, the CBS is assumed to have near to perfect CSI by periodic updates. This may lead to extra overheads in the network but for functionality of a centralized CRN, a near to perfect CSI is a must. 

CBS is the main entity responsible for scheduling user access, especially if there are multiple SUs. Further, 
CBS is expected to constantly monitor PU activity and then ask SU to immediately adjust its transmit power according to PU demand\footnote{In this study, we have assumed that there would be “constant monitoring by CBS”. However, there needs to be a periodicity of monitoring. Such period can be same as the arrival rate of Poisson distribution or further optimized considering the power efficiency.}. 
In addition, the thermal additive white Gaussian noise (AWGN) in the network is assumed to have circularly symmetric complex Gaussian distribution with zero mean and variance as $\sigma^2$, i.e., $\mathcal{CN}(0, \sigma^2)$. Finally, to improve the readability of this paper, the most frequently  used symbols are described in Table \ref{ds}. 

\noindent\begin{table}[!ht] 
\centering
\caption{Notations.}\label{ds}
\noindent\begin{tabularx}{\columnwidth}{>{\raggedright\arraybackslash}X >{\raggedright\arraybackslash} X }
\hline 
Symbol & Description  \\ \hline
c & Capacity\\ 
$\gamma_p$ & SINR at PU-Rx \\
$\gamma_s$ & SINR at SU-Rx \\
$\alpha$ &  Continuous random variable for PU-Rx SINR  \\
$\alpha_k$ & Discrete  random variable for PU-Rx SINR, where $k=1,2,\dots,\infty$ \\
$\psi$ & Interference plus noise\\
$\lambda_p$ & Poisson rate parameter\\
$\lambda_{xx}$ & Channel rate parameter (exponential rate) with subscript $x$ can be $s$ or $p$ implying SU or PU\\
$\sigma^2$ & AWGN variance\\
$P_{rx}$ & Received power at PU-Rx\\
$p$ &  Peak  transmit power\\
$x$ & Dummy variable\\
\hline
\end{tabularx}
\end{table}
\section{Modelling Interference Temperature from Capacity distribution} \label{sec3}

In this section, we will derive the interference power threshold from the variable traffic demand distribution considering a CRN system as described in the previous section\footnote{Note that the interference power threshold and interference temperature (IT) are used interchangeably throughout the paper.}. First of all, the data traffic distribution or the variable capacity distribution is discrete in nature, and therefore has been modelled by different available discrete distribution's  \cite{disc1,disc2,disc3,disc4,disc5}. However, among those discrete distributions, Poisson distribution becomes a very strong candidate as it has been used in telecommunications since the advent of computer networks, and with proper selection of parameters can be made to fit to most network traffic models \cite{mandelbrot1965self,frost1994traffic}. Also, since the traffic  capacity demand change over time is a discrete quantity, and the occurrences of the traffic demand events are independent from each other, the applicability of  Poisson distribution to a network traffic model is further strengthened. Apart from that, in different scenarios and conditions, Poisson distribution has been shown  to match, and model the traffic data in a network  \cite{lopez2013time,ding2017spectrum,disc2,disc3}.

Ideally, one can use any of the available continuous or discrete distributions, but  considering the close applicability of discrete distributions and usage for capacity demand modeling,  Poisson distribution is found to be the best model to represent the PU capacity demand\cite{disc1,disc2,disc3,disc4,disc5}. In addition, the insights provided in this study by modelling capacity demand by any of the available  distributions will remain the same. The traffic/capacity demand assuming  Poisson distribution with rate parameter as $\lambda_p$  is given as,
  \begin{equation}
       P_c(x_k)= \frac{\lambda_p^{x_k}}{x_k! }e^{-\lambda_p} ,  \  \forall \  \lambda_p >0,\ x_{k} \in \{0,1,2,3,\dots, \infty \}.
 \end{equation}
In Poisson distribution, this  $\lambda_p $ is also the  mean parameter, which in our case  represents the mean capacity value.

Now, we will use this Poisson distributed capacity demand to find SINR distribution, and afterwards from that SINR distribution, IT distribution is determined. These statistical random variable transformation are done step by step by applying well known CDF method \cite{papoulis2002probability,miller2012probability}. To begin with, the relationship between instantaneous capacity and SINR is given as,

\begin{equation}
    c=\log (1+\gamma_p), \ \forall \ \gamma_p \ge 0, \nonumber
\end{equation}
where  $\gamma_p$ represents the instantaneous SINR at PU-Rx\footnote{SINR is usually denoted by $\gamma$, but in this paper   $\gamma_p$ and $\gamma_s$ are used to easily distinguish between the SINR at PU-Rx and SINR at SU-Rx respectively.}and $c$ represents the instantaneous capacity  in (nats/s)/Hz. Using transformation of random variables, the probability mass function (PMF) of SINR  at PU-Rx is given as,  
\begin{equation}
\begin{split}
    P (\gamma_p)&=P(c) \big |_{\gamma_p=e^c-1}=  \frac{e^{-\lambda_p}\lambda_p^{c_k}}{c_k! } \Bigg |_{c=\log(1+\gamma_p)}.\\
       \end{split}
   \end{equation}        

This transformed discrete distribution can be easily expressed as a continuous distribution \cite{pishro2016introduction} as follows,        
    \begin{equation} \label{sinrNoTrunc}
      f_{\gamma_p} (x)=\mathlarger{\sum}_{x_{k} \in \mathbb{R}} \frac{e^{-\lambda_p}\lambda_p^{\log(x_k+1)}}{\log(x_k+1)! } \delta(x - x_k). \ \ \forall \  x \ge 0,  
    \end{equation}
where $\delta(x)$ is a dirac delta function. This transformation from discrete to continuous random variable will save a lot of effort in  computations involving mixed random variable distributions (continuous and discrete distributions) to derive the network performance expressions. However, this simple discrete random transformation needs more careful inspection.

\begin{figure}[t]
\centering
\includegraphics{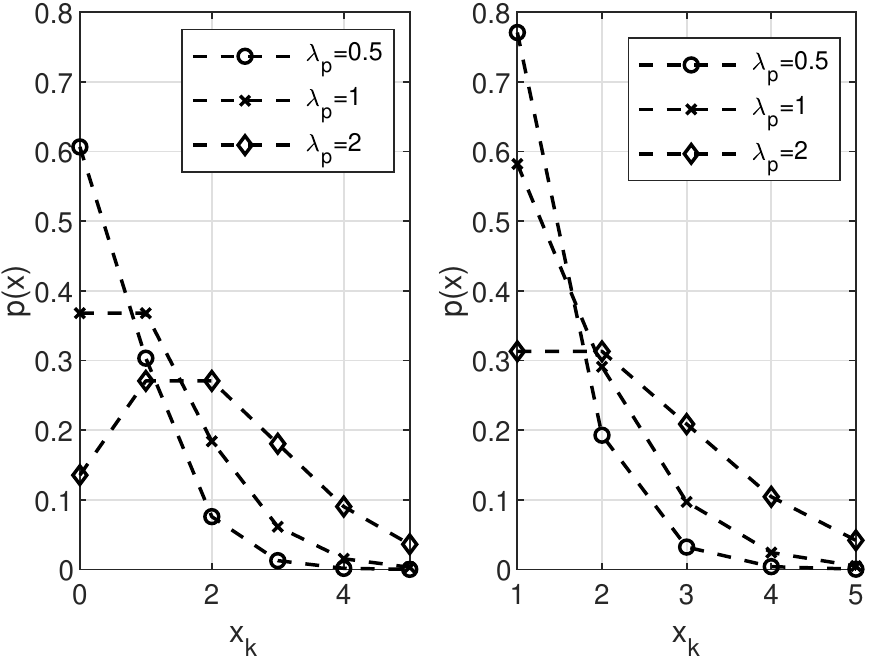}
 \caption{Comparison of normal Poisson distribution (left) with zero truncated Poisson distribution (right) for different rate parameter of $\lambda_p=0.5,1,2$. Here the line plot is used for better visualization.}
\label{TrunPois}
\end{figure}

The SINR at the PU-Rx is the ratio of the received signal  power from PU-Tx to the interference plus noise power, that is $\gamma_p =P_{PU}^{Rx}/\psi$, where $\psi$ represents the interference  plus noise power. Therefore,

\begin{equation}\label{pssi}
    \psi=\frac{g_{pp}p}{\gamma_p}, \forall \ \gamma_p=\{0,1,2, \dots,\infty\} \in \mathbb{R},  
\end{equation}
where $p$ is the peak transmit power\footnote{The peak transmit power of SU and PU is assumed to be the same.}. For $\gamma_p=0$ in \eqref{pssi},  $\psi$ would be undefined, which will lead to  an undefined  data distribution. Therefore, to overcome this fallacy, we can either begin by truncating the PU capacity distribution from Poisson to zero-truncated Poisson distribution \cite{griffin1992distribution,singh1978characterization,cohen1960estimating},  or we can  truncate the distribution of SINR ($\gamma_p$) itself, which will have the range of $\gamma_p=\{1,2, \dots,\infty\}$, that is $ \gamma_p \in \mathbb{R}^+$. Fig. \ref{TrunPois} shows a case of truncating capacity distribution from  general  Poisson distribution to a zero-truncated Poisson distribution. One can observe the increase in the probabilities of all the samples after truncation with shape remaining unchanged, for example at $x_k=2, \lambda_p=2$, the probability is $P(x)=0.2707$ and at the same parameters in zero-truncated Poisson distribution the probability is $P(x)=0.313$. The increase in the probability is  due to the shrinking  of the sample space. 

In our case, we will proceed with the SINR truncation method. Since,   $ f_{\gamma_p}  (x) =1$ for $x \ge 0$ and  $ f_{\gamma_p}  (x)=0$ for $x <0$,  hence we can write $ f_{\gamma_p} (x )$ for the case of $x \ge 0$ as,
\begin{equation}\label{truncSnr}
\begin{split}
       f_{\gamma_p} (x)\big|_{x\ge0} & =  f_{\gamma_p}  (x)\big |_{x=0} + f_{\gamma_p}(x) \big |_{x > 0},  \\
        f_{\gamma_p}  (x ) \big |_{x > 0} &=1-e^{-\lambda_p x}.
\end{split}
\end{equation}

Hence, the  SINR  distribution with support region of $ \gamma_p \in \mathbb{R}^+$ will be then given by normalizing \eqref{sinrNoTrunc} by \eqref{truncSnr}, that is,
\begin{equation}\label{SINR}
\begin{split}
     f_{\gamma_p}  (x )&=\mathlarger{\sum}_{x_{k}\in  \mathbb{R}^+} \frac{e^{-\lambda_p}\lambda_p^{\log(1+x_k)}}{ (1-e^{-\lambda_p})\log(1+ x_k)! } \delta (x-x_k), \ \  x >0.
\end{split}
   \end{equation}

 \begin{figure}[t]
\centering
\includegraphics[width=1\linewidth]{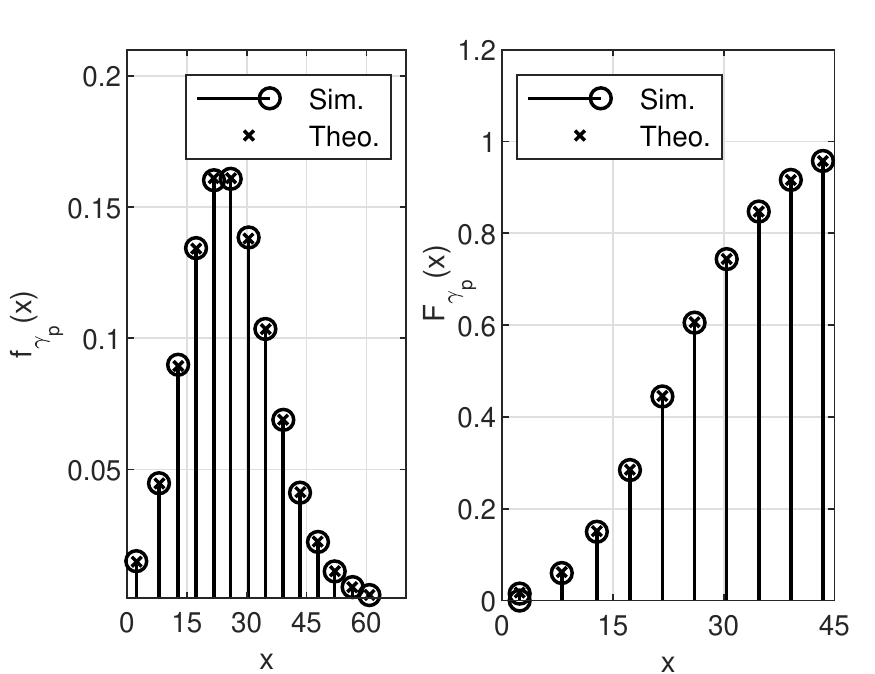}
 \caption{Simulation and theoretical PDF and CDF of SINR for $\lambda_p=6$.}
\label{SINR_poiss}
\end{figure}  

Fig. \ref{SINR_poiss} compares the PDF and CDF generated from the above theoretical SINR expression with simulations results\footnote{In simulations, the primarily utilized MATLAB functions are: ``poissrnd" for generating random numbers from Poisson distribution, ``ecdf" for empirical cumulative distribution function and ``histogram" for generating probabilistic plots for the simulation data.}, which are in complete agreement with each other. Once the SINR distribution at PU is known,  the  interference plus noise distribution will be then just a ratio of two random variables, that is,
\begin{equation} 
    \begin{split}
     \psi=\frac{g_{pp}p}{\gamma_p}=\bigg(\frac{g_{pp}}{\gamma_p}\bigg) p. 
     \end{split}
\end{equation}

On careful observation, the numerator of this equation has a random variable $g_{pp}$, which is exponentially distributed channel power, while the denominator $\gamma_p$ is the SINR random variable, whose distribution was derived in \eqref{SINR} using statistical transformations from capacity demand. The correlation between these two random variable is zero, hence, they  are both independent. The only dependency between these random variable is $\psi$, which has to be dynamically adjusted. In other words, the PU is allowed to transmit the data at maximum power $p$, so as to have high PU network capacity, while  the SINR derived from the capacity is made dependent on the wireless channel conditions ($g_{pp}$) through IT ($\psi$). This in turn forces  the SUs  to adapt their transmit power accordingly with the dynamic IT set at PU-Rx.

Let $\frac{P_{PU}^{Rx}}{\gamma_p}=\frac{\beta}{\alpha}$, using the CDF method for ratio of two random variables \cite{miller2012probability,pishro2016introduction}, the PDF is then given as,
\begin{equation} 
\begin{split}
        f_\psi (x)&=\int_0^{\infty}\alpha \cdot f_{P_{rx},\gamma_p}(x\alpha ,\alpha) d \alpha, \\
       \end{split}
\end{equation} 
Since, $\beta$ and $\alpha$ are independent random variables, therefore,
 \begin{equation} \label{7}
\begin{split}       
       f_\psi (x) &=\int_0^{\infty} \alpha \cdot f_{P_{rx}}(x \alpha)\  f_{\gamma_p }(\alpha)\ d \alpha,  \forall \ \alpha >0. 
       \end{split}
\end{equation} 

As the channel is assumed to be Rayleigh distributed, the channel power distribution will be a scaled exponential distribution, 
\begin{equation} \label{scaledExp}
    f_{P_{rx}}(x)=\frac{\lambda_{pp}}{p} e^{-\lambda_{pp} x/p}, 
\end{equation}
where $\lambda_{pp}$ is the channel rate parameter between PU-Tx and PU-Rx, while $p$ is the peak PU transmit power.

Substituting \eqref{scaledExp} and \eqref{SINR} in \eqref{7}, 
       \begin{equation}
    \begin{split} 
       f_\psi (x) &=\int_0^{\infty}\alpha  \bigg (\frac{  \lambda_{pp}}{p} e^{-\lambda_{pp} \alpha x/p}\bigg)\bigg (\mathlarger{\sum}_{\alpha_{k}\in  \mathbb{R}^+}  \frac{e^{-\lambda_p}}{1-e^{-\lambda_p}}\\
      &  ~~~~~~~~~~~~~~~~~~~~~~~~\times \frac{\lambda_p^{\log(1+\alpha_k)}}{\log(1+\alpha_k)! } \delta (\alpha-\alpha_k)\bigg)  d\alpha ,\\
        &=\frac{\lambda_{pp}}{p} \mathlarger{\sum}_{\alpha_{k}\in  \mathbb{R}^+} \frac{e^{-\lambda_p}\lambda_p^{\log(1+\alpha_k)}}{(1-e^{-\lambda_p})\log(1+\alpha_k)! } \int_0^{\infty} \alpha e^{- \frac{\lambda_{pp}\alpha_k x}{p}}\\
        &~~~~~~~~~~~~~~~~~~~~~~~~~~~~~~~~~~~~~~~~~~~\times \delta (\alpha-\alpha_k) d\alpha,
        \end{split}
\end{equation}
Using the property of delta function, $\int_{\alpha -\epsilon}^{\alpha +\epsilon}f(t) \delta(t-\alpha) dt= f(\alpha),\ \epsilon >0$, the expressions reduces to,
\begin{equation}\label{ITpdf}
    f_\psi(x)=   \frac{\lambda_{pp}}{p}\mathlarger{\sum}_{\alpha_{k}\in  \mathbb{R}^+} \frac{ e^{-\lambda_p}\lambda_p^{\log(1+\alpha_k)}}{(1-e^{-\lambda_p})\log(1+\alpha_k)! }  \alpha_k e^{-\lambda_{pp} \alpha_k x/p}.
\end{equation}

Also, the CDF of  interference plus noise is found by integrating the PDF as,
\begin{equation} 
    \begin{split}
        F_\psi(x)&= \int_0^{x} \frac{\lambda_{pp}}{p}\mathlarger{\sum}_{\alpha_{k}\in  \mathbb{R}^+} \frac{ e^{-\lambda_p}\lambda_p^{\log(1+\alpha_k)}\alpha_k e^{-\lambda_{pp} \alpha_k x/p}}{(1-e^{-\lambda_p})\log(1+\alpha_k)! }    d x,\\ 
        &= \frac{\lambda_{pp}}{p} \frac{ e^{-\lambda_p}}{1-e^{-\lambda_p}}\mathlarger{\sum}_{\alpha_{k}\in  \mathbb{R}^+} \frac{\lambda_p^{\log(1+\alpha_k)}\alpha_k}{\log(1+\alpha_k)! }  \\
        & ~~~~~~~~~~~~~~~~~~~~~~~~~\times \int_0^{x} e^{-\lambda_{pp} \alpha_k x/p} d x.
          \end{split}
\end{equation}

On further evaluation,  the  final CDF expression comes out to be 
\begin{equation} \label{IT_cdf}
    \begin{split}
        F_\psi(x)&=\mathlarger{\sum}_{\alpha_{k}\in  \mathbb{R}^+} \frac{e^{-\lambda_p}\lambda_p^{\log(1+\alpha_k)}}{(1-e^{-\lambda_p})\log(1+\alpha_k)! } \big [1- e^{-\lambda_{pp} \alpha_k x/p} \big ].
    \end{split}
\end{equation}

 \begin{figure}[t]
\centering
\includegraphics[width=3.45in,height=2.7in]{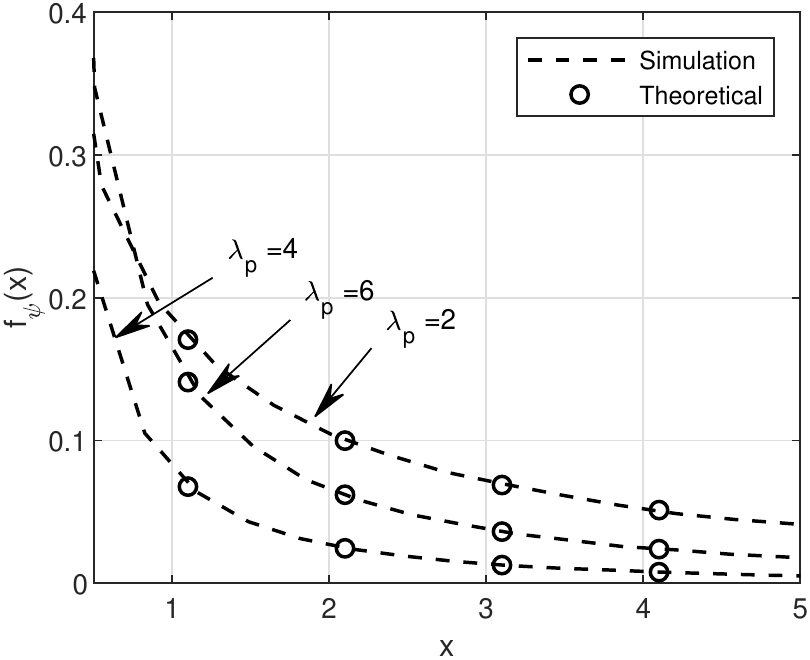}
 \caption{Simulation and theoretical PDF of interference plus noise for different values of $\lambda_p=2,4,6$.}
\label{NI1}
\end{figure}

Fig. \ref{NI1} shows the match between the simulation and theoretical result for the PDF derived in  \eqref{ITpdf} for  different $\lambda_p$. Since, the noise is assumed to be Gaussian distribution with zero mean and variance as $\sigma^2$, i.e., $\mathcal{CN}(0, \sigma^2)$, the interference plus noise can be considered as total interference threshold or IT with a constant noise variance $\sigma^2$ included in it. In simpler terms,  $\psi$ can be regarded as total interference threshold. The proof that \eqref{ITpdf} is a valid PDF is given in Appendix \ref{app1}. 

So far the first step of deriving the dynamic distribution of IT for PU-Rx from the variable network traffic demand (Poisson distributed) has been found, the next step is to evaluate the SU network performance metrics by deriving and analyzing outage probability and mean capacity distributions, which will be the topic of discussion in the next section.

\section{Secondary user Outage Probability and mean capacity}  \label{sec4}

In this section, we will derive the outage probability and mean capacity of SU assuming peak power adaptation at SU-Tx. The SU transmit power with peak power adaptation\cite{PeakVsAverage} is given as,
\begin{equation} \label{min2}
    P_{SU}^{Tx}=\text{min} \ \bigg( \frac{\psi}{g_{sp}},p \bigg),
\end{equation}
where $p$ is the peak transmit power, $g_{sp}$ is the channel between SU-Tx to PU-Rx and $\psi$ is the total IT. To make the mathematical notation's simpler, we will use $P_{tx}$ and $P_{rx}$ for transmit and receiver power at SU, rather than $P_{SU}^{Tx}$ and $P_{SU}^{Rx}$. Therefore, the SINR at SU-Rx is,
\begin{equation}\label{gamma}
    \gamma_s =\frac{P_{tx} g_{ss}}{ p g_{ps} +\sigma^2}=\frac{P_{rx}}{p g_{ps} +\sigma^2} ,
    \end{equation}
where  $g_{ss}$ is the channel power gain between SU-Tx and SU-Rx, while $g_{ps}$ is the  channel power gain between PU-Tx and SU-Rx, and $p$ is the peak power at PU. The mean capacity expression for SU is therefore,

\begin{equation}
    \begin{split}
        \bar{C} &= B \int_0^{\infty} \log(1+x)\  f_{\gamma_s} (x) dx, \\
           \end{split}
\end{equation} 
where $f_{\gamma_s} (x)$ is the PDF of SINR at SU-Rx, and $B$ is the bandwidth. Using integration by parts \cite{suraweera2010capacity}, the expression can be written as,
\begin{equation}
    \begin{split}
       \bar{C}   &= \int_0^{\infty} \frac{1-F_{\gamma_s}(x)}{1+x}dx,
    \end{split}
\end{equation}
where $B$ is assumed to be 1 Hz and $F_{{\gamma}_s}(x)$ is the CDF of SINR or the outage probability. Therefore, to evaluate the mean capacity, we need to find the  outage probability at SU.

Here also, we will do step by step statistical transformation of random variables to  derive $\gamma_s$ from $P_{tx}$, and then finally  the outage probability $F_{\gamma_s}$ and capacity $C$. To determine $P_{tx}$ as given in \eqref{min2}, let  $t = \frac{\psi}{g_{sp}}=\frac{u}{v}$. Then, the CDF of $t$ will be given as,
\begin{equation}
\begin{split}
       F_T(x)&= \Pr \bigg( \frac{u}{v} <x \bigg)=\Pr \big( u < v x, \ v>0 \big),\\
        & =\int_0^{\infty} \int_0^{v x}  f_{\psi, g_{sp}} (uv)du \  dv.
     \end{split}
   \end{equation}    

Since $\psi$ and $g_{sp}$ are independent random variable, therefore,
\begin{equation}
\begin{split} 
      F_T(x)  &= \int_0^{\infty} \int_0^{v x}f_\psi (u) du  f_{g_{sp}}(v) dv,\\
        &= \int_0^{\infty} F_\psi(v x) f_{g_{sp}}(v) dv.
\end{split}
   \end{equation}
where $f_{g_{sp}}(v)$ is the PDF of the exponential channel power gain, and $F_{\psi}(x)$ is given in \eqref{IT_cdf}. On substituting these terms,

\begin{equation}
    \begin{split}
        F_T(x)&=  \int_0^{\infty}\mathlarger{\sum}_{\alpha_{k} \in \mathbb{R}^+} \frac{e^{-\lambda_p}\lambda_p^{\log(1+\alpha_k)} \big [1- e^{\frac{-\lambda_{pp} \alpha_k v x}{p}} \big ] }{(1-e^{-\lambda_p})\log(1+\alpha_k)! }   \\
    &~~~~~~~~~~~~~~~~~~~~~~~~~~~~~~~~~~~~~ \times \lambda_{sp} e^{-\lambda_{sp}v} dv,\\    
             & = \frac{\lambda_{sp} e^{-\lambda_p} }{1-e^{-\lambda_p}} \mathlarger{\sum}_{\alpha_k \in \mathbb{R}^+} \frac{\lambda_p^{\log(1+\alpha_k)}}{\log(1+\alpha_k)!} \bigg [  \int_0^{\infty}  e^{-\lambda_{sp}v} dv \\          
        &~~~~~~~~~~~~~~~~~~~~~~~~ - \int_0^{\infty}  e^{- \big( \frac{\lambda_{pp}\alpha_k x}{p} +\lambda_{sp} \big) v} dv \bigg ].
               \end{split}
\end{equation}
which finally reduces to,
\begin{equation}\label{F_t}
\begin{split}
   F_T(x) &= \frac{ e^{-\lambda_p} }{1-e^{-\lambda_p}}\mathlarger{\sum}_{\alpha_k \in \mathbb{R}^+}                     \frac{\lambda_p^{\log(1+\alpha_k)}}{\log(1+\alpha_k)!}\  \bigg( \frac{\eta \alpha_k x}{\eta \alpha_k x +p}\bigg),
\end{split}
\end{equation}
where $\eta = \frac{\lambda_{pp}}{\lambda_{sp}}$. The PDF of $t = \frac{\psi}{g_{sp}}$ will be then given as,

\begin{equation}\label{f_t}
    f_T(x)=\frac{ e^{-\lambda_p} }{(1-e^{-\lambda_p})}\mathlarger{\sum}_{\alpha_k \in \mathbb{R}^+}                     \frac{\lambda^{\log(1+\alpha_k)}}{\log(1+\alpha_k)!}\  \bigg[ \frac{\eta \alpha_k p}{(\eta \alpha_k x +p)^2}\bigg].
\end{equation}

Thus, the distribution of $P_{tx}$ as a minimum function of the constant $p$ and random variable $t$ is as follows,
\begin{equation}\label{F_ptx}
\begin{split}
    P_{tx} &= \text{min} \ \bigg( \frac{\psi}{g_{sp}},p \bigg) =\text{min} \ ( t,p ), \nonumber\\
       F_{P_{tx}}(x) &= F_T(x) +F_p(x) - F_T(x)F_p(x),\\
         &= F_T(x) \ [ 1- H(x-p)] + H(x-p),
\end{split}
\end{equation}
where the constant $p$ is expressed as a random variable with CDF as a Heaviside function $H(x-p)$ and PDF as a Dirac delta function $\delta(x-p)$ \cite{kachroo2018impact,pishro2016introduction}. Correspondingly, the PDF would be given as,
\begin{equation}\label{f_ptx}
    \begin{split}
        f_{P_{tx}}(x) &=f_T(x) [1-H(x-p)]+\delta(x-p)\\
        &=~~~~~~~~~~~~~~~~~~~~-F_T(x)\delta(x-p),
    \end{split}
\end{equation}
where $F_T(x)$ and $f_T(x)$  are given in \eqref{F_t} and \eqref{f_t}.

From this given $P_{tx}$ distribution, the next step is to  determine the received power at SU-Rx, that is, $P_{rx}= P_{tx} g_{ss}$. This expression is a product of two random variables, one of which is exponential random variable $g_{ss}$, and the other one is $P_{tx}$. Let $P_{rx}= P_{tx} g_{ss}$ be written as $P_{rx}=v_1 v_2$ where $v_1=g_{ss}$ and $v_2=P_{tx}$. Therefore,
\begin{equation}\label{tt}
    \begin{split}
        F_{P_{rx}} (x) &= \text{Pr}(v_1 v_2 \le x)\\
        &= \int_0^\infty \int_0^{\frac{x}{v_2}} f_{g_{ss}}\bigg( v_1\le \frac{x}{v_2}\bigg) f_{P_{tx}}(v_2) d v_2,\\
        &= \int_0^\infty F_{g_{ss}}( x/v_2)f_{P_{tx}}(v_2) d v_2,\\
        &= \int_0^\infty [1- e^{-\lambda_{ss} x/v_2}]\ f_{P_{tx}}(v_2) d v_2,\\
        &= 1- \int_0^\infty e^{-\lambda_{ss} x/v_2}\ f_{P_{tx}}(v_2) d v_2. \\
            \end{split}
\end{equation}  

On substituting \eqref{F_t},  \eqref{f_t} and \eqref{f_ptx} in  \eqref{tt},
        \begin{equation}
    \begin{split}
      F_{P_{rx}} (x)  &= 1- \int_0^\infty e^{-\lambda_{ss} x/v_2} \big[  f_{t}(v_2) [1-H(v_2-p)]  \\
        &~~~~~~~~~~~~~~+\delta(v_2-p) -F_{{t}}(v_2) \delta(v_2-p)\big ] dv_2,\\
         &= 1- \int_0^{\infty}     \frac{e^{-\lambda_{ss} x/v_2}  e^{-\lambda_p} }{(1-e^{-\lambda_p})}\mathlarger{\sum}_{\alpha_k \in \mathbb{R}^+} \frac{\lambda_p^{\log(1+\alpha_k)}}{\log(1+\alpha_k)!}\\
        &~~~~~~\times \bigg[ \frac{\eta \alpha_k p}{(\eta \alpha_k v_2 +p)^2}\bigg] d v_2 + \int_p^{\infty}  \frac{ e^{-\frac{\lambda_{ss} x}{v_2}}  e^{-\lambda_p} }{(1-e^{-\lambda_p})}\\
        &\times \mathlarger{\sum}_{\alpha_k \in \mathbb{R}^+} \frac{\lambda_p^{\log(1+\alpha_k)}}{\log(1+\alpha_k)!}\bigg[ \frac{\eta \alpha_k p}{(\eta \alpha_k v_2 +p)^2}\bigg] dv_2  - e^{-\frac{\lambda_{ss} x}{p}} \\
        &~~~~~~~~~~~~~~ + \int_0^{\infty} e^{-\lambda_{ss} x/v_2} F_{P_{tx}}(v_2) \delta(v_2-p) dv_2,\\
           \end{split}
\end{equation}
which finally reduces down to, 
\begin{equation} \label{F_prx}
    \begin{split}
           F_{P_{rx}} (x)  &= 1- e^{-\lambda_{ss} x/p} + \frac {e^{-\lambda_p}} {1-e^{-\lambda_p}}\mathlarger{\sum}_{\alpha_k \in \mathbb{R}^+} \frac{\lambda^{\log(1+\alpha_k)}}{\log(1+\alpha_k)!} \\
           &~~~~~~~ \times \frac{\eta \alpha_k \lambda_{ss} x}{p} e^{\frac{\eta \alpha_k \lambda_{ss} x}{p}} \Gamma \bigg( 0,\frac{\lambda_{ss} (\eta \alpha_k +1) x}{p} \bigg), 
    \end{split}
\end{equation}
where $\Gamma(a,x)$ is an incomplete Gamma function \cite{stein}, which is defined as,
\begin{equation}
    \Gamma(a,x)=\int_{a}^{\infty} t^{a-1}e^{-t} dt,  \ \forall \ a>0,\ x \ge 0. \nonumber
\end{equation} 

Finally, from this distribution of $P_{rx}$, the distribution of SINR at SU-Rx will be the  ratio of two independent random variables $P_{rx}$ and $p g_{ps} +\sigma^2$, which can be easily found out by using the same CDF method. That is,
\begin{equation}
 \gamma_s =\frac{P_{rx}}{ p g_{ps} +\sigma^2}.
\end{equation}
Let $\frac{P_{rx}}{ p g_{ps} +\sigma^2} =\frac{u}{v}$, therefore,

\begin{equation}\label{F_gamma}
\begin{split}
      F_{\gamma_s}(x)&=\text{Pr} \bigg( \frac{u}{v} \le x\bigg),\\ 
            &= \int_{\sigma^2}^{\infty} \int_0^{v x} f_{P_{rx}}(u\le v x) f_v(v)\ du \ dv,\\
            &= \int_{\sigma^2}^{\infty} \bigg( \int_0^{v x} f_{P_{rx}}(u \le v x)\ du \bigg) \ f_v(v) dv,\\
            &= \int_{\sigma^2}^{\infty} F_{P_{rx}}(v x) \ f_v(v) dv.
\end{split}
\end{equation}

Since, $v=p g_{ps} +\sigma^2$ is a scaled and shifted exponential distribution, while $F_{P_{rx}}(v x)$ was found in \eqref{F_prx}. Therefore, 
\begin{equation}
    \begin{split}
        F_{\gamma_s}(x) &= \int_{\sigma^2}^{\infty} \bigg [ 1- e^{-\frac{\lambda_{ss} x}{p}} + \frac {e^{-\lambda_p}} {1-e^{-\lambda_p}}\mathlarger{\sum}_{\alpha_k \in \mathbb{R}^+} \frac{\lambda_p^{\log(1+\alpha_k)}}{\log(1+\alpha_k)!} \\
           &~~~\times \frac{\eta \alpha_k \lambda_{ss}  v x}{p} e^{\frac{\eta \alpha_k \lambda_{ss} v x}{p}} \Gamma \bigg(0, \frac{\lambda_{ss} (\eta \alpha_k +1) vx}{p} \bigg)   \bigg ]\\
           &~~~~~~~~~~~~~~~~~~~~~~~~~~~~~~~~~\times \frac{\lambda_{ps}}{p} e^{-\frac{\lambda_{ps}}{p} (v-\sigma^2)} dv,\\
         &=  \int_{\sigma^2}^{\infty} \frac{\lambda_{ps}}{p} e^{-\frac{\lambda_{ps}}{p} (v-\sigma^2)} dv -  \int_{\sigma^2}^{\infty} \frac{\lambda_{ps}}{p}  e^{-\frac{\lambda_{ss} x}{p}}    \\
         & ~~~~~~\times e^{-\frac{\lambda_{ps}}{p} (v-\sigma^2)} dv + \int_{\sigma^2}^{\infty} \frac {e^{-\lambda_p}} {1-e^{-\lambda_p}}   \\
         & ~~~~~\times \mathlarger{\sum}_{\alpha_k \in \mathbb{R}^+} \frac{\lambda_p^{\log(1+\alpha_k)}}{\log(1+\alpha_k)!} \frac{\eta \alpha_k \lambda_{ss} v x}{p} e^{\frac{\eta \alpha_k \lambda_{ss} v x}{p}}\\
         &  ~~~~~\times   \Gamma \bigg( 0,\frac{\lambda_{ss} (\eta \alpha_k +1) v x}{p} \bigg) \frac{\lambda_{ps}}{p}e^{-\frac{\lambda_{ps}}{p} (v-\sigma^2)} dv, \\
                   \end{split}
\end{equation}
which can be expressed as,
\begin{equation}
    F_{\gamma_s}(z)  = I_1- I_2 - I_3.
\end{equation}

Evaluating these integrals individually,
  \begin{equation}
     I_1 =\int_{\sigma^2}^{\infty} \frac{\lambda_{ps}}{p} e^{-\frac{\lambda_{ps}}{p} (v-\sigma^2)} dv=1. 
 \end{equation}
 
 Here $I_1$ is the PDF that is integrated over its full range resulting the value to be 1. $I_2$ on the other hand is evaluated as,
 
\begin{equation}
    \begin{split}
        I_2 &= \int_{\sigma^2}^{\infty}   e^{-\lambda_{ss} x/p}   \frac{\lambda_{ps}}{p} e^{-\frac{\lambda_{ps}}{p} (v-\sigma^2)} dv \\
        &~~~~~~= \frac{\lambda_{ps}e^{\lambda_{ps} \sigma^2}}{p} \int_{\sigma^2}^{\infty} e^{-\frac{\lambda_{ss} x + \lambda_{ps}}{p}v} dv,\\
            \end{split}
\end{equation}
which on further evaluation yields,
\begin{equation}
    \begin{split}
        I_2&=  \frac{\lambda_{ps}}{\lambda_{ps} + \lambda_{ss} x} e^{-\frac{\lambda_{ss} \sigma^2 x}{p}}.
    \end{split}
\end{equation}

Lastly, evaluating $I_3$,
\begin{equation}
    \begin{split}
        I_3 &=  \frac {e^{-\lambda_p}\lambda_{ps}} { (e^{-\lambda_p}-1)p}  e^{\frac{\lambda_{ps}}{p} \sigma^2}  \mathlarger{\sum}_{\alpha_k \in \mathbb{R}^+} \frac{\lambda_p^{\log(1+\alpha_k)}}{\log(1+\alpha_k)!}\frac{\eta \alpha_k \lambda_{ss} x}{p} \\
         & ~~~\times \int_{\sigma^2}^{\infty}  v e^{\frac{\eta \alpha_k \lambda_{ss} v x}{p}} e^{-\frac{\lambda_{ps}}{p} v} \Gamma \bigg( 0,\frac{\lambda_{ss} (\eta \alpha_k +1) v x}{p} \bigg) d v, 
         \end{split}
 \end{equation}
 
 Which on further evaluation yields,
         \begin{equation}
            \begin{split}
     I_3&= \mathlarger{\sum}_{\alpha_k \in \mathbb{R}^+} \frac{ \eta \lambda_{ps} \lambda_{ss}\alpha_k x e^{-\lambda_p}(e^{-\lambda_p}-1)^{-1}}{(\lambda_{ps}- \eta \alpha_k \lambda_{ss} x)^2 (\lambda_{ps}+ \lambda_{ss}x) }    \\
  &~~\times  \frac{\lambda_p^{\log(1+\alpha_k)}}{\log(1+\alpha_k)!} \bigg[(\lambda_{ps} - \eta \alpha_k \lambda_{ss} x) e^{-\lambda_{ss}\sigma^2 x/p}  \\
  & ~~+(\lambda_{ps}+\lambda_{ss}x) \bigg\{e^{\eta \alpha_k \lambda_{ss}\sigma^2 x/p} \bigg( 1+ \frac{(\lambda_{ps}-\eta \alpha_k \lambda_{ss})\sigma^2}{p}\bigg)\\
  &~~~~~~~~~~~~~~~\times \Gamma \bigg(0,\frac{(\eta \alpha_k +1) \lambda_{ss} \sigma^2 x}{p}\bigg) -e^{\lambda_{ps}\sigma^2/p}\\
  & ~~~~~~~~~~~~~~~~~~~~~~~~~~~~\times \Gamma \bigg(0,\frac{(\lambda_{ps} + \lambda_{ss} x) \sigma^2}{p} \bigg) \bigg \} \bigg].
\end{split}
 \end{equation}
 
 Therefore, on substituting $I_1$, $I_2$ and $I_3$, $F_{\gamma_s}(x)$ or the outage probability  will be given as  in \eqref{outage}, and the proof that it is valid distribution is given in Appendix \ref{app2}. Furthermore, substituting \eqref{outage} in the mean capacity expression that is given  as,
\begin{equation}\label{meanCapacity}
  \bar{C}= \int_0^{\infty} \frac{1-F_{\gamma_s}(x)}{1+x}dx .
\end{equation}
will result in \eqref{capacity}. Unfortunately,  there is no closed form solution for the second integral  ($I_4$) of \eqref{capacity}, which therefore needs to be evaluated numerically. Thus far, we have derived the outage probability and mean capacity of a SU in a CRN network irrespective of  their operable region. However, there can be a case where the peak power $p$ is very high, so that during the peak power adaptation at SU-Tx, the final transmit power ($P_{SU}^{Tx}$)  will always be $\frac{\psi}{g_{sp}}$. This region is also known as high power region \cite{PeakVsAverage}, and we will analyze this region for more deeper insights. In the next section, we will derive the corresponding performance expressions for such  region.
 
 \begin{figure*}[htbp]
\begin{equation}\label{outage}
\begin{split}
     F_{\gamma_s}(x)  &= 1 -\frac{\lambda_{ps}e^{-\frac{\lambda_{ss} \sigma^2 x}{p}}}{\lambda_{ps} + \lambda_{ss} x} + \mathlarger{\sum}_{\alpha_k \in \mathbb{R}^+} \frac{ \eta \lambda_{ps} \lambda_{ss}\alpha_k x e^{-\lambda_p} (1-e^{-\lambda_p})^{-1}}{(\lambda_{ps}- \eta \alpha_k \lambda_{ss} x)^2 (\lambda_{ps}+ \lambda_{ss}x) }   \frac{\lambda_p^{\log(1+\alpha_k)}}{\log(1+\alpha_k)!}  
  \bigg[ (\lambda_{ps} - \eta \alpha_k \lambda_{ss} x) e^{-\lambda_{ss}\sigma^2 x/p}+(\lambda_{ps}\\
  & +\lambda_{ss}x) \bigg\{e^{\eta \alpha_k \lambda_{ss}\sigma^2 x/p}
 \bigg( 1+ \frac{(\lambda_{ps}-\eta \alpha_k \lambda_{ss})\sigma^2}{p}\bigg)\Gamma\bigg(0,\frac{(\eta \alpha_k +1) \lambda_{ss} \sigma^2 x}{p}\bigg)
 -e^{\lambda_{ps}\sigma^2/p} \Gamma\bigg(0,\frac{(\lambda_{ps} + \lambda_{ss} x) \sigma^2}{p} \bigg) \bigg \} \bigg]
    \end{split}
\end{equation}
\vspace*{0.3cm}
\\
\rule[0.2cm]{1\textwidth}{0.017cm}
\end{figure*}

\begin{figure*}[ht]
\begin{equation}\label{capacity}
\begin{split}
    \bar{ C} &= \int_0^\infty \frac{\lambda_{ps}e^{-\frac{\lambda_{ss} \sigma^2 x}{p}}}{ (\lambda_{ps} + \lambda_{ss} x)(1+x)}  d x - \int_0^\infty  \mathlarger{\sum}_{\alpha_k \in \mathbb{R}^+} \bigg( \frac{ \eta \lambda_{ps} \lambda_{ss}\alpha_k x e^{-\lambda_p}}{(\lambda_{ps}- \eta \alpha_k \lambda_{ss} x)^2 (\lambda_{ps}+ \lambda_{ss}x) (1-e^{-\lambda_p}) (1+x)} \frac{\lambda_p^{\log(1+\alpha_k)}}{\log(1+\alpha_k)!}  
   \\
  & \times \bigg[ (\lambda_{ps} - \eta \alpha_k \lambda_{ss} x) e^{-\lambda_{ss}\sigma^2 x/p}dx +(\lambda_{ps}+\lambda_{ss}x) \bigg\{e^{\eta \alpha_k \lambda_{ss}\sigma^2 x /p}
 \bigg( 1+ \frac{(\lambda_{ps}-\eta \alpha_k \lambda_{ss})\sigma^2}{p}\bigg)\Gamma\bigg(0,\frac{(\eta \alpha_k +1) \lambda_{ss} \sigma^2 x}{p}\bigg)\\
 &~~~~~~~~~~~~~~~~~~~~~~~~~~~~~~~~~~~~~~~~~~~~~~~~~~~~~~~~~~~~~~~~~~~~~~~~~~~~~~~~~~~~~~~~ -e^{\lambda_{ps}\sigma^2/p} \Gamma \bigg (0,\frac{(\lambda_{ps} + \lambda_{ss} x) \sigma^2}{p} \bigg) \bigg \} \bigg] \bigg) dx\\
  \bar{ C} &=\frac{\lambda_{ps}}{\lambda_{ss}-\lambda_{ps}}\bigg[ e^{\lambda_{ps} \sigma^2/p} \Gamma \bigg(0,\frac{\lambda_{ps} \sigma^2}{p}\bigg)- e^{\lambda_{ss} \sigma^2/p} \Gamma\bigg(0,\frac{\lambda_{ss} \sigma^2}{p}\bigg)\bigg] + I_4\\
   \end{split}
\end{equation}
\vspace*{0.2cm}
\\
\rule[0.2cm]{1\textwidth}{0.017cm} 
\end{figure*}

\section{Outage probability and mean Capacity in high power region}  \label{sec5}
In high power region, when $p>> \frac{\psi}{g_{sp}}$ at SU-Tx, the SU transmit power assuming peak power adaptation is given as,

\begin{equation} \label{min}
    P_{SU}^{Tx}=\text{min} \ \bigg( \frac{\psi}{g_{sp}},p \bigg)=\frac{\psi}{g_{sp}}=t,
\end{equation}
which reduces the probability distribution of $P_{tx}$ to $t(x)$ as given in \eqref{f_ptx}. Therefore, the probability distribution of $P_{rx}$ as $t =\frac{u}{v}$ is as follows,

\begin{equation}
    \begin{split}
        F_{P_{rx}}(x)&=\int_0^\infty F_{g_{ss}}( x/v)f_{P_{tx}}(v) d v,\\
        &= \int_0^\infty [1- e^{-\frac{\lambda_{ss} x}{v}}]\ \frac{ e^{-\lambda_p} }{1-e^{-\lambda_p}}\mathlarger{\sum}_{\alpha_k \in \mathbb{R}^+}                     \frac{\lambda_p^{\log(1+\alpha_k)}}{\log(1+\alpha_k)!}\\
        & ~~~~~~~~~~~~~~~~~~~~~~~~~~~~~~~~~~~\times \bigg[ \frac{\eta \alpha_k p}{(\eta \alpha_k x +p)^2}\bigg] d v,\\
                  \end{split}
\end{equation}
which simplifies down to
\begin{equation}
\begin{split}
   F_{P_{rx}}(x)&= 1- \frac{ e^{-\lambda_p} }{1-e^{-\lambda_p}}\mathlarger{\sum}_{\alpha_k \in \mathbb{R}^+}\frac{\lambda_p^{\log(1+\alpha_k)}}{\log(1+\alpha_k)!}\bigg[ 1-\frac{\eta \alpha_k \lambda_{ss}}{p} \\
   & ~~~~~~~~~~~~~~~~~~\times x e^{\lambda_{ss} \eta \alpha_k x/p}\cdot\Gamma \bigg(0,\frac{\lambda_{ss}\eta \alpha_k x}{p}\bigg) \bigg].
\end{split}
 \end{equation}
 
Thus, the outage probability will be,
\begin{equation}
    \begin{split}
       F_{\gamma_{s}}(x) &= \int_{\sigma^2}^{\infty} F_{P_{rx}}(v x) f_v(v) dv,\\
       &= \int_{\sigma^2}^{\infty} \bigg [ 1- \frac{ e^{-\lambda_p} }{1-e^{-\lambda_p}}\mathlarger{\sum}_{\alpha_k \in \mathbb{R}^+}\frac{\lambda_p^{\log(1+\alpha_k)}}{\log(1+\alpha_k)!} \\
   & ~\times \bigg( 1-\frac{\eta \alpha_k \lambda_{ss}  v x e^{\lambda_{ss} \eta \alpha_k v x/p}}{p} \Gamma\bigg(0, \frac{\lambda_{ss}\eta \alpha_k x v}{p}\bigg)\bigg)\bigg]\\
   &~~~~~~~~~~~~~~~\times  \frac{\lambda_{ps}}{p} e^{-\frac{\lambda_{ps}}{p} (v-\sigma^2)} dv,\\
         \end{split}
\end{equation} 
   
  \begin{equation}
    \begin{split}
  &=1-\frac{ e^{-\lambda_p} }{1-e^{-\lambda_p}} \mathlarger{\sum}_{\alpha_k \in \mathbb{R}^+}                     \frac{\lambda_p^{\log(1+\alpha_k)}}{\log(1+\alpha_k)!} - \frac{ e^{-\lambda_p} }{1-e^{-\lambda_p}}\\
  & ~~~~\times \mathlarger{\sum}_{\alpha_k \in \mathbb{R}^+} \frac{\lambda_p^{\log(1+\alpha_k)}\eta \alpha_k \lambda_{ss}x}{\log(1+\alpha_k)!}\frac{1}{(\lambda_{ps}-\eta \alpha_k \lambda_{ss}x)^2} \\
  &~~~~~~~~~ \times \bigg[ \lambda_{ps} -\eta \alpha_k \lambda_{ss}x +\lambda_{ps}e^{\lambda_{ps}\sigma^2/p}\Gamma \bigg(0, \frac{\lambda_{ps} \sigma^2}{p}\bigg) \bigg], \\
     \end{split}
\end{equation}
which on further evaluation reduces to,
   \begin{equation} \label{highpowerOutage}
    \begin{split}
     F_{\gamma_{s}}(x) &= \mathlarger{\sum}_{\alpha_k \in \mathbb{R}^+} \frac{\lambda_p^{\log(1+\alpha_k)}\eta \alpha_k \lambda_{ss}x}{\log(1+\alpha_k)!}\frac{1}{(\lambda_{ps}-\eta \alpha_k \lambda_{ss}x)^2} \\
   &\times \bigg[ \lambda_{ps} -\eta \alpha_k \lambda_{ss}x -\lambda_{ps}e^{\lambda_{ps}\sigma^2/p} \Gamma \bigg(0, \frac{\lambda_{ps} \sigma^2}{p}\bigg) \\
    &  +\lambda_{ps}e^{\eta \alpha_k \lambda_{ss}x \sigma^2/p}\bigg( 1+\frac{(\lambda_{ps}-\eta \alpha_k \lambda_{ss}x) \sigma^2}{p}\bigg)  \\
     &~~~~~~~~~~~~~~~~~~~~~~~~~~~~ \times  \Gamma \bigg (0,\frac{\eta \alpha_k \lambda_{ss}x \sigma^2}{p}\bigg) \bigg].
       \end{split}
\end{equation}

Finally, to determine the mean capacity, we will substitute \eqref{highpowerOutage} in \eqref{meanCapacity}. Unfortunately, the expression also doesn't have a closed form solution. Therefore, the expression  has to be evaluated numerically. 

In next section, we will validate and discuss all these derived expressions with simulation results in detail. 

\section{Simulation Results and discussion} \label{sec6}
In this section, we will compare the analytical expression derived in the previous sections by comparing them with Monte Carlo Simulations in MATLAB\textsuperscript \textregistered, and numerical evaluations in MATHEMATICA\textsuperscript \textregistered \footnote{It is highly recommended to use MATHEMATICA for numerical evaluation involving high powers as it provides excellent numerical precision at these high power values.}. Furthermore, instantaneous and mean capacity, and outage probability of a dynamic IT based CRN, are  compared with a fixed IT based CRN to show the advantages of setting a dynamic IT over fixed IT. 
The different distance dependent rate parameters used in the simulation were selected for illustrative purposes, but depending upon different scenario's (environment and distance), different values can be used. Nevertheless, the selection of parameters are inconsequential to the insights provided by choosing any set of rate parameters. For Rayleigh fading channels, the mean values were selected as  $E(g_{sp})=1/ \lambda_{sp}=2$, $E(g_{ps})=1/ \lambda_{ps}=3.3$, $E(g_{ss})=1/ \lambda_{ss}=5$ and $E(g_{pp})=1/ \lambda_{pp}=4$, while the peak power was chosen  depending on the analysis and case in hand. Also, the noise power was set to be $\sigma^2=1$. These different  parameters  are also summarized in the Table \ref{nk}.

\setlength{\extrarowheight}{1pt}
\begin{table}[!ht]
\centering 
\caption{Different simulation parameters.}
\label{nk}
\begin{tabular}{|c|c|}
\hline
Parameter & Value \\ \hline
$E[g_{sp}]$ & 2 \\ \hline
$E[g_{ps}]$ & 3.3 \\ \hline
$E[g_{ss}]$ & 5 \\ \hline
$E[g_{pp}]$ & 4 \\ \hline
$\sigma^2$ & 1 \\ \hline
p & -10,0,10 dB (depending on the given case) \\ \hline
$\lambda_p$ & 1,2...,6 (depending on the given case) \\ \hline
\end{tabular}
\end{table}

\begin{figure}[!t]
\centering
\includegraphics{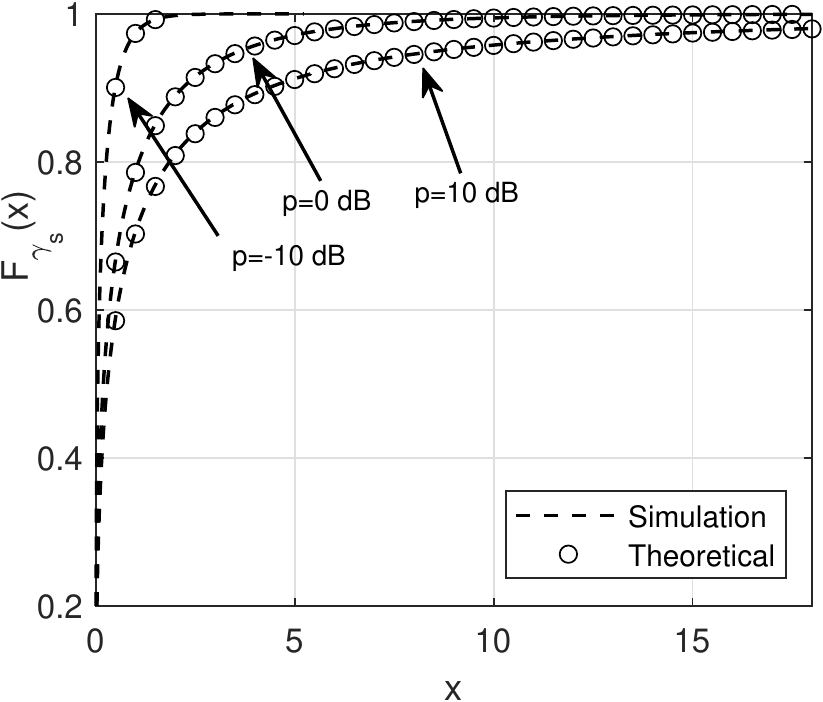}
 \caption{Simulation and theoretical outage probability in general region of SU at different peak power values of $p=-10,0, +10$ dB.}
\label{OP}
\end{figure}

First, we start with the simulation of the outage probability expression given in expression \eqref{outage}. In this case,  the capacity demand (Poisson distribution) at PU  is fixed  at $\lambda_p=2$, while peak power is chosen to be $-10$, $0$ and $+10$ dB. Fig. \ref{OP} shows the result of comparison between the simulation and theoretical expression that are in total agreement. Intuitively, high transmit power will better accommodate the capacity/data traffic demand than the low transmit power, which is reflected in the outage probability plot of Fig. \ref{OP} with less outage values. As an example, one can observe that at every SINR value in Fig. \ref{OP}, the outage probability is higher for low peak power value as compared to  higher peak power value.
 
\begin{figure}[!t]
\centering
\includegraphics{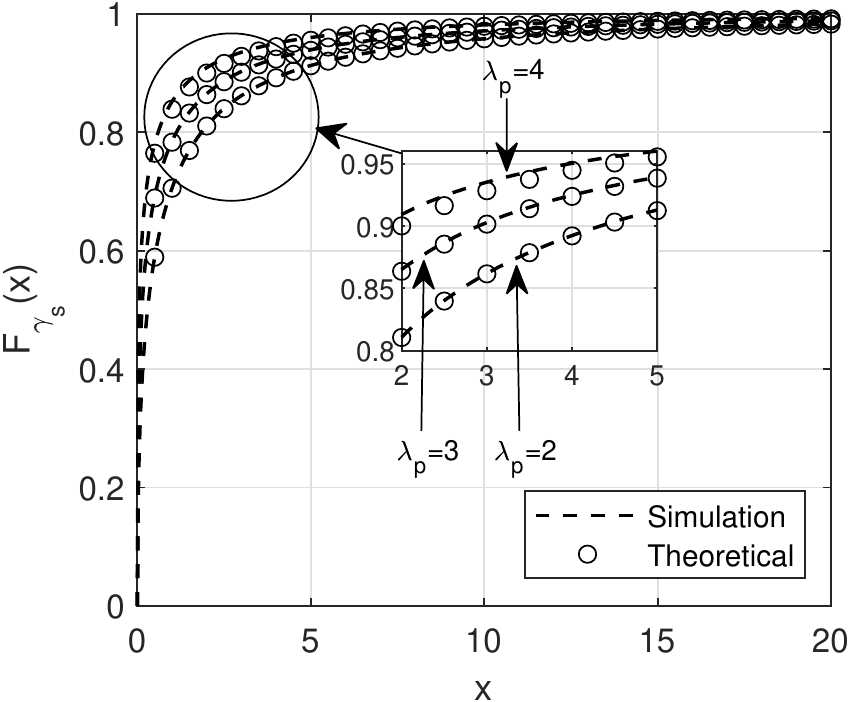}
 \caption{Simulation and theoretical outage probability in general power region of SU at  $\lambda_p=2,3,4$}.
\label{OP2}
\end{figure}

Next, we fix the peak transmit power at $10$ dB and analyze the effect of changing capacity/data traffic demand of PU on outage probability. The changing capacity demand is reflected in the Poisson rate parameter  and the  values  selected in this scenario are  $\lambda_p=2,3$ and $4$. It can be intuitively inferred that a high capacity demand from a PU should set the dynamic IT very tight, thereby restricting the transmit power for SU. Correspondingly, a low capacity demand from a PU should relax the IT,  allowing the SU to opportunistically  increase their transmit power. This phenomenon can be easily seen from Fig. \ref{OP2}, where for any SINR value, the outage in case of $\lambda_p=4$ is more than  $\lambda_p=3$, and  that is even more than the case of $\lambda_p=2$. In case of high power region, the expression given in \eqref{highpowerOutage} will yield similar inferable results.

   \begin{figure}[!t]
\centering
\includegraphics{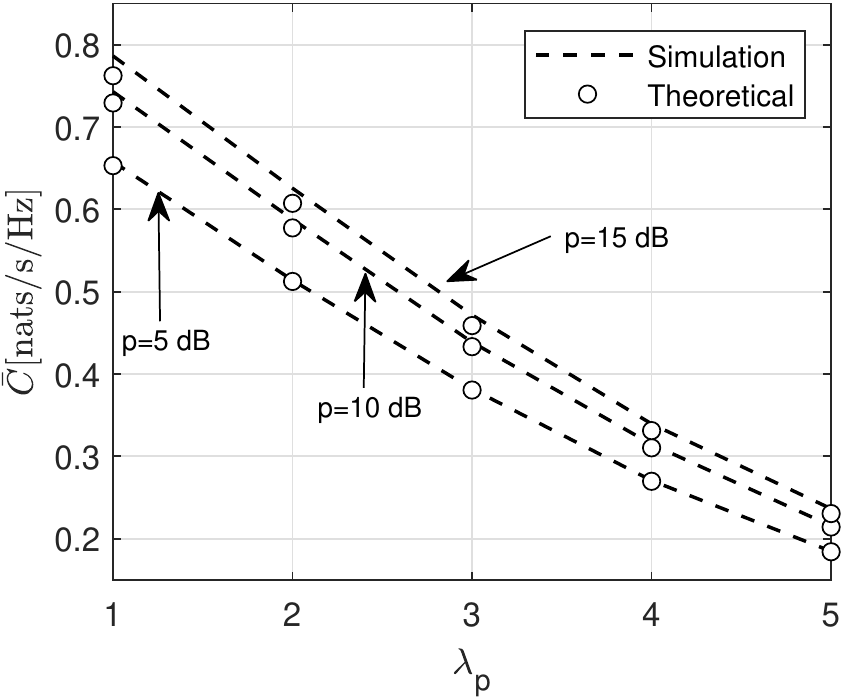}
 \caption{Simulation and theoretical mean capacity of SU at  $p=5,10,15$ dB,  while varying the $\lambda_p$ from 1 to 5.}
\label{MeanCap2}
\end{figure}

Moving forward, first the effect of varying $\lambda_p$ (capacity parameter) at  fixed SU transmit power levels, and then the  effect of changing peak power at various fixed  $\lambda_p$  will be  evaluated and analyzed  for the mean capacity expression  generated for SU in \eqref{capacity}.  For the first case, we select three peak power levels of $p=5,10,15$ dB, while  $\lambda_p$ is varied from $1$ to $5$. As one may expect, high peak transmit power will allow the SU to have higher capacity as compared to low peak transmit power, however, as the $\lambda_p$ increases (high PU capacity demand), the dynamic IT that will be set by CBS for SU will become more tighter. This tight IT will therefore limit the SU transmit power, ultimately resulting to a lower mean capacity. This interesting phenomenon can be easily observed in  Fig. \ref{MeanCap2} in which the simulation and theoretical results are plotted.

 \begin{figure}[!t]
\centering
\includegraphics{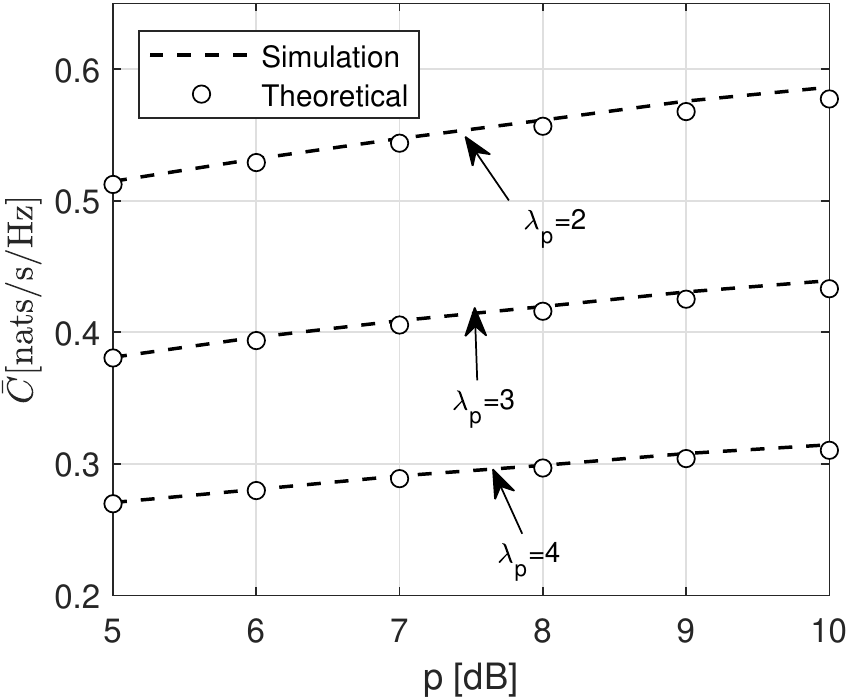}
 \caption{Simulation and theoretical mean capacity of SU at different values of $\lambda_p=2,3,4$ while varying the peak power from $5$ to $10$ dB.}
\label{MeanCap1}
\end{figure}

 \begin{figure}[!t]
\centering
\includegraphics{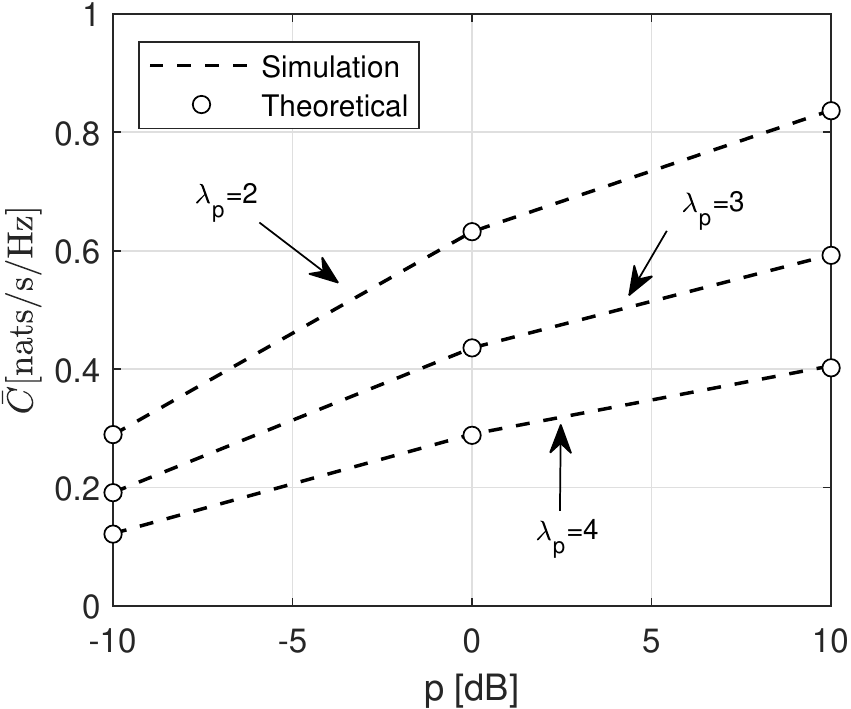}
 \caption{Simulation and theoretical mean capacity of SU at different values of $\lambda_p=2,3,4$ while varying the peak power from $-10$ to $10$ dB in high power region.}
\label{cap33}
\end{figure}

 \begin{figure}[!ht]
\centering
\includegraphics{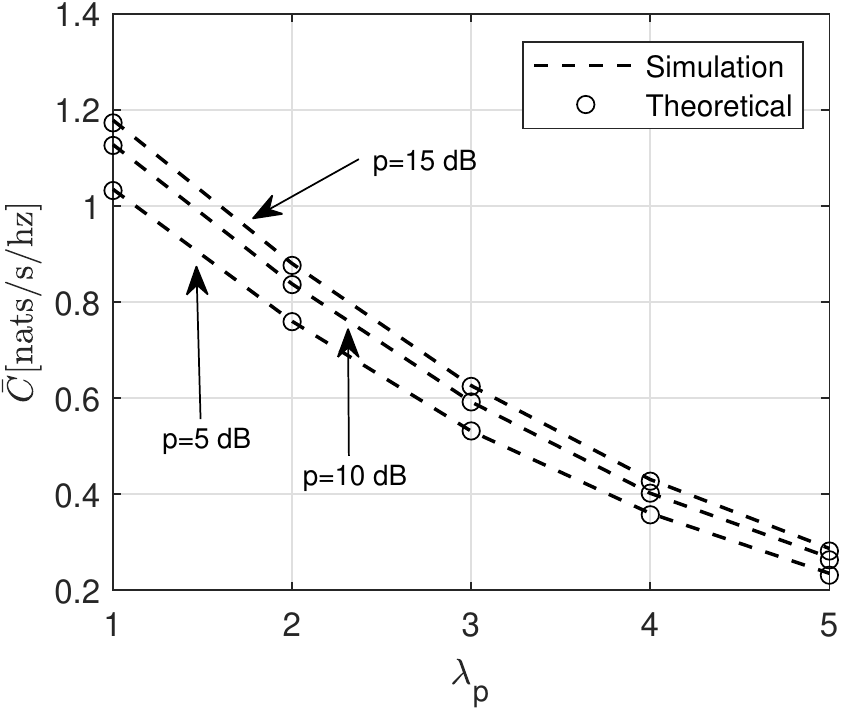}
 \caption{Simulation and theoretical mean capacity of SU at  $p=5,10,15$ dB,  while varying  $\lambda_p$ from 1 to 5 in high power region.}
\label{cap34}
\end{figure}

  \begin{figure}[!ht]
\centering
\includegraphics{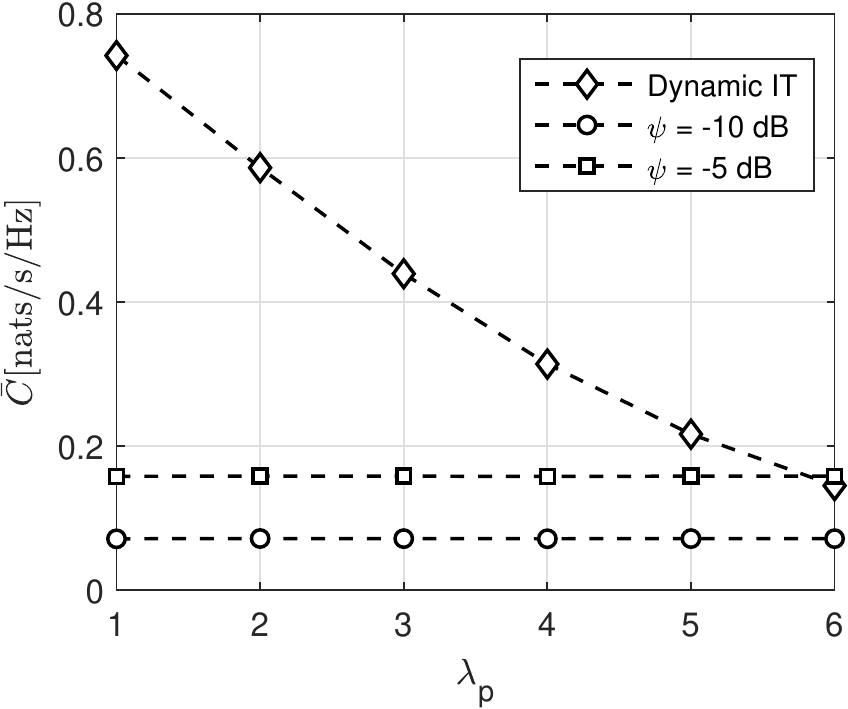}
 \caption{ Comparison of mean capacity performance with dynamic IT and fixed IT of  $\psi=-10$ and $-5$ dB.}
\label{compcap}
\end{figure}

In the next step,  the peak power  is varied from $5$ to $10$ dB and  the mean capacity is evaluated at three different values of $\lambda_p$ as $2,3$ and $4$. As expected, increasing  peak power of SUs at relaxed dynamic IT (small $\lambda_p$) will increase the SU mean capacity than at  low SU peak power with  high $\lambda_p$. This mechanism is due to the  dynamic setting of  IT,  which is governed by the PU capacity changes (capacity demand). Fig. \ref{MeanCap1} shows this resulting plots, where the network dynamics (traffic data demand) is reflected in the setting of dynamic IT and which in turn gets reflected in the mean capacity. The same is  also inferred in the high power region for which the mean capacity expressions are  evaluated numerically.  Fig. \ref{cap33} and Fig. \ref{cap34} show the corresponding mean capacity plots for SU in the high power region.

Finally, we will simulate and compare the case of using dynamic IT with the fixed IT values set at $-10$ and $-5$ dB values for mean and instantaneous capacity performance, with outage probability. One important point to note here is that these set values for fixed IT are pre-chosen to be very tight, and  represent the worst case scenario. In this  case, the peak power is fixed at $10$ dB, and   $\lambda_p$ is varied from 1 to 6 to generate the dynamic IT, and thereby the mean capacity. Fig. \ref{compcap} shows the resulting plot of such a case, and  it can be observed that the mean capacity achieved  by a SU with dynamic IT is higher than that with the  fixed IT one as it takes care of the capacity variation of PU. In other words, a smaller value of $\lambda_p$ will relax the IT, thereby allowing SU to have high transmit power and therefore, high mean capacity. On the other hand, a high value of $\lambda_p$ will reflect a tight IT value for SU, thus limiting the SU transmit power which finally results in low mean capacity.

\begin{figure}[t]
\centering
\includegraphics{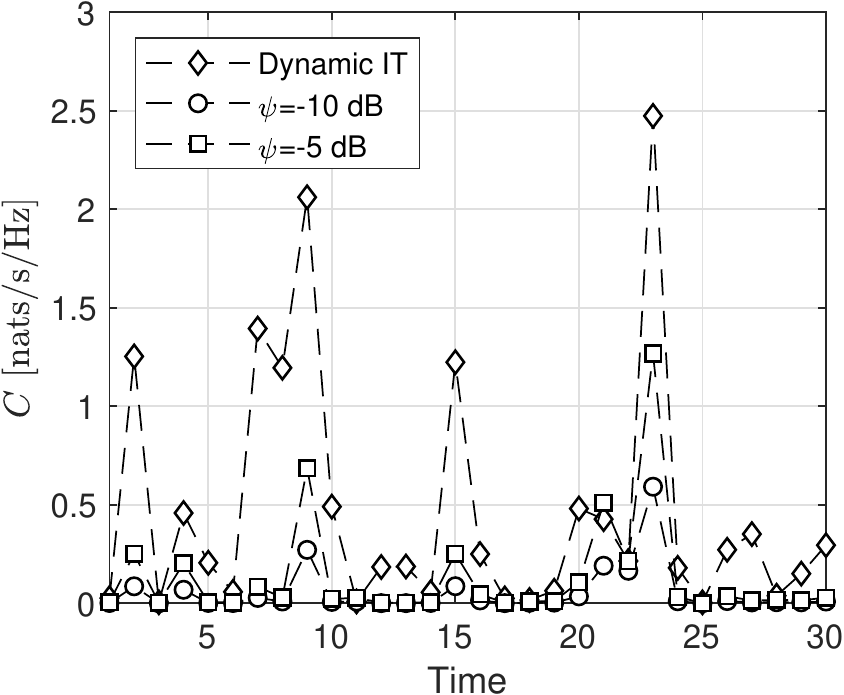}
 \caption{Comparison of instantaneous capacity performance with dynamic IT and fixed IT of  $\psi=-10$ and $-5$ dB.}
\label{compcap31}
\end{figure}

In the second case, we will compare the simulated instantaneous capacity performance with dynamic IT, and   with the fixed  IT values set at $-10$ and $-5$ dB. Fig. \ref{compcap31} shows the result for 30 time flops, and it can be clearly observed that setting dynamic IT leads to better instantaneous network capacity than the fixed IT case with the same reasoning as in the previous case.

\begin{figure}[t]
\centering
\includegraphics[width=1\linewidth]{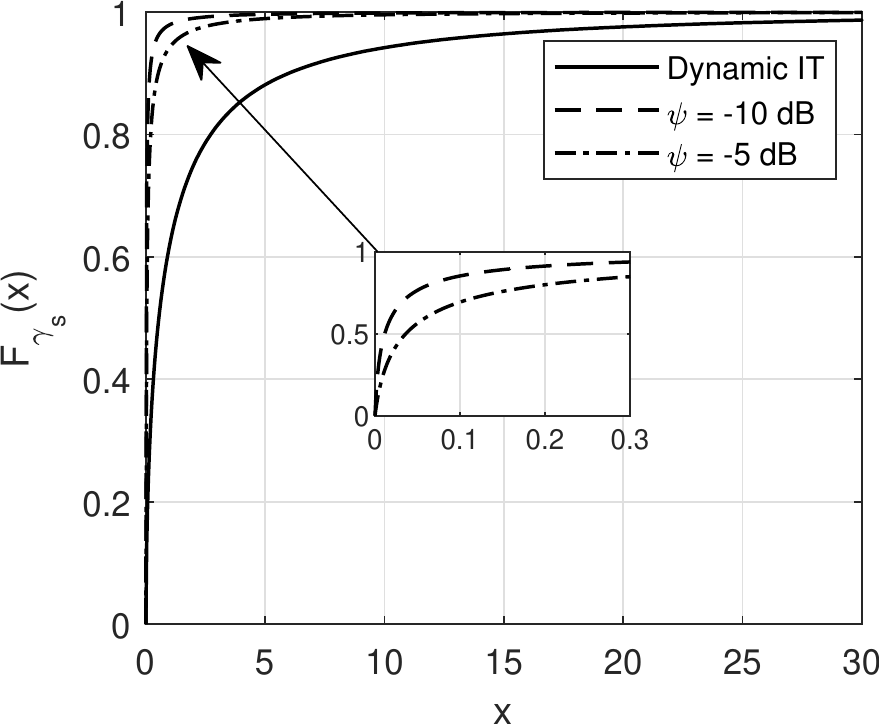}
 \caption{Comparison of outage probability with dynamic IT with fixed IT kept at $\psi=-10$ and $-5$ dB.}
\label{outDyn}
\end{figure}

In the last case, we will compare the simulated outage probability of SU with respect to dynamic IT and fixed IT kept at $-10$ and $-5$ dB with $\lambda_p$ set at 1. Please note that the  effect of varying $\lambda_p$ on outage probability is already shown in Fig. \ref{OP2}. As expected with the setting of dynamic IT,  the outage probability would be less than the fixed IT case at any given SINR value, which can be easily observed from Fig. \ref{outDyn}. Also, one can observe that the outage probability in case of stricter IT, which is kept at -10 dB, is more than the case with fixed IT of -5 dB. 
Therefore, the  positive effects of setting IT as a dynamic value can be easily seen on the  outage probability, and on the  mean and instantaneous capacity of SU, which could be easily leveraged in designing spectrum sharing systems for future.

\section{Conclusion and Future work}
In this work, we have statistically modelled the dynamic interference temperature or interference power threshold  from the variable capacity demand of PU in a cognitive radio system. The PU capacity demand variation over  time was  assumed to follow Poisson distribution, and  consequently, using statistical transformations of random variables, the distribution of SINR, and finally IT distribution was found and validated. Theoretical expressions for outage probability and mean capacity for SU in the general power region, and in the high power region were  derived and verified with simulation results. Finally, we analyzed the effect of utilizing a dynamic interference power threshold on the mean and instantaneous capacity, and on the outage probability. We found  that   the dynamic IT  substantially improves the  network performance as compared to a fixed IT based cognitive radio system. If obtained, the asymptotic analysis of capacity and outage probability expressions in  general, and in high power region could provide valuable insights. These are considered as future work.


\ifCLASSOPTIONcaptionsoff
  \newpage
\fi

\appendices
\section{Proof of valid PDF }\label{app1}
For \eqref{ITpdf} to be a valid PDF, it  should satisfy these two necessary conditions,
\begin{itemize}
    \item $f_\psi(x) \ge 0,\ \forall x.$
    \item $\int_{-\infty}^{+\infty} f_\psi(x)dx=1 $.
\end{itemize}
where $f_\psi(x)$ is given as,

\begin{equation}
    f_\psi(x)=   \frac{\lambda_{pp}}{p}\mathlarger{\sum}_{\alpha_{k}\in  \mathbb{R}^+} \frac{ e^{-\lambda_p}\lambda_p^{\log(1+\alpha_{k})}}{(1-e^{-\lambda_p})\log(1+\alpha_{k})! }  \alpha_{k} e^{-\lambda_{pp} \alpha_{k} x/p}. \nonumber
\end{equation}
The first property is easy to prove.  The second property can be proved as follows,
\begin{equation}
    \begin{split}
        & \int_{0}^{\infty}   \frac{\lambda_{pp}}{p}\mathlarger{\sum}_{\alpha_{k}\in  \mathbb{R}^+} \frac{ e^{-\lambda_p}\lambda_p^{\log(1+\alpha_{k})}\alpha_{k}}{(1-e^{-\lambda_p})\log(1+\alpha_{k})! }   e^{-\lambda_{pp} \alpha_{k} x/p} dx,\\ \nonumber
        &=\frac{\lambda_{pp}}{p} \mathlarger{\sum}_{\alpha_{k}\in  \mathbb{R}^+} \frac{ e^{-\lambda_p}\lambda_p^{\log(1+\alpha_{k})}\alpha_{k}}{(1-e^{-\lambda_p})\log(1+\alpha_{k})! }   \int_{0}^{\infty} e^{-\lambda_{pp} \alpha_{k} x/p} dx,\\
         \end{split}
\end{equation}
\begin{equation}
    \begin{split}
        &= \frac{ e^{-\lambda_p}}  {1-e^{-\lambda_p}}\mathlarger{\sum}_{\alpha_{k}\in  \mathbb{R}^+} \frac{\lambda_p^{\log(1+\alpha_{k})}}{\log(1+\alpha_{k})! } .\\ \nonumber
           \end{split}
\end{equation}

Using Taylor series, the expression reduces to,
\begin{equation}
    \begin{split}
        \int_{0}^{\infty} f_\psi(x)dx &=\frac{e^{-\lambda_p}}{1-e^{-\lambda_p}} (e^{\lambda_p}-1),\\ \nonumber
       \int_{0}^{\infty} f_\psi(x)dx &=1.
    \end{split}
\end{equation}
Hence, it is a valid PDF.

\section{Proof of valid CDF }\label{app2}
For \eqref{outage} to be a valid CDF it  should satisfy these two necessary conditions,
\begin{itemize}
    \item $\lim\limits_{x\to -\infty}  F_\gamma(x)=0$.
    \item $\lim\limits_{x\to \infty}  F_\gamma(x)=1 $.
\end{itemize}
where $ F_\gamma(x)$ is given in \eqref{outage}. Now, at  $x\to 0$ the second term in \eqref{outage} will reduces to, 
\begin{equation}
    \begin{split}
        &\frac{\lambda_{ps}}{\lambda_{ps} + \lambda_{ss} x} e^{-\frac{\lambda_{ss} \sigma^2 x}{p}}=1.
    \end{split}
\end{equation}
while the third term is $0$. Therefore $F_\gamma(x)=0$ as $x\to 0$.  Evaluating at $x \to \infty$, 
\begin{equation}
    \begin{split}
        e^{-\frac{\lambda_{ss} \sigma^2 x}{p}}&=0,\\
        \Gamma(0,\infty)=0.
    \end{split}
\end{equation}

Therefore, the second  and third term  in \eqref{outage} becomes $0$,  and  $F_\gamma(x)$ will be $1$. Hence the expression is a valid CDF.

\section*{Acknowledgment}
The authors would like to thank the editor for handling the manuscript, and reviewers for their valuable feedback and suggestions. We would also like to thank, Md. Zobaer Islam, for his help in  proofreading the manuscript. In addition, this work was supported by the National Science Foundation under Grant Number 1923295.
\bibliographystyle{IEEEtran}

\bibliography{biblog}

\begin{IEEEbiography}[{\includegraphics[width=1in,height=1.25in,keepaspectratio]{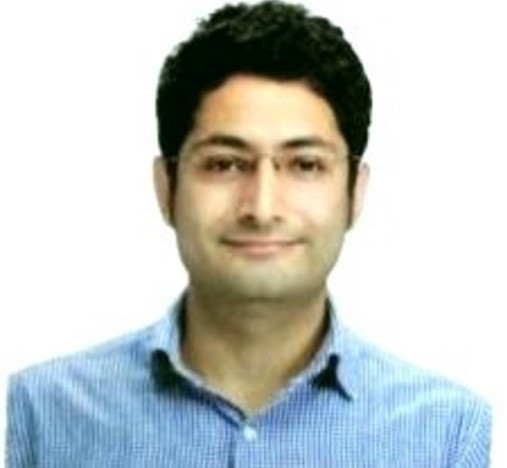}}]{Amit Kachroo (S'14)}  received his B.Tech degree in Electronics and Communications Engineering from NIT, Srinagar, India (2005-2009), and M.Sc degree from Istanbul Sehir University, Istanbul, Turkey (2015-2017) respectively. From 2010 to 2014, he worked as a project engineer with Nokia Networks, India on GSM, WCDMA, LTE, and RF technologies. In 2017, he joined Oklahoma State University, Stillwater, USA  to pursue  his Ph.D. degree in Electrical and Computer Engineering.  In summer of 2019, he worked  as a software intern at Tensilica R\&D group of Cadence Design Systems, California, USA in  mmWave radar technology.  His research interest are in statistical modeling of wireless channels, mmWave channel modeling, statistical learning, machine learning, and cognitive radios.  He  also serves as a reviewer in  IEEE Communications Magazine, IEEE Transactions on Wireless Communications, IEEE JSAC, IEEE Access, and IEEE VTC Conference among others.  
\end{IEEEbiography}

\begin{IEEEbiography}[{\includegraphics[width=1in,height=1.25in,clip]{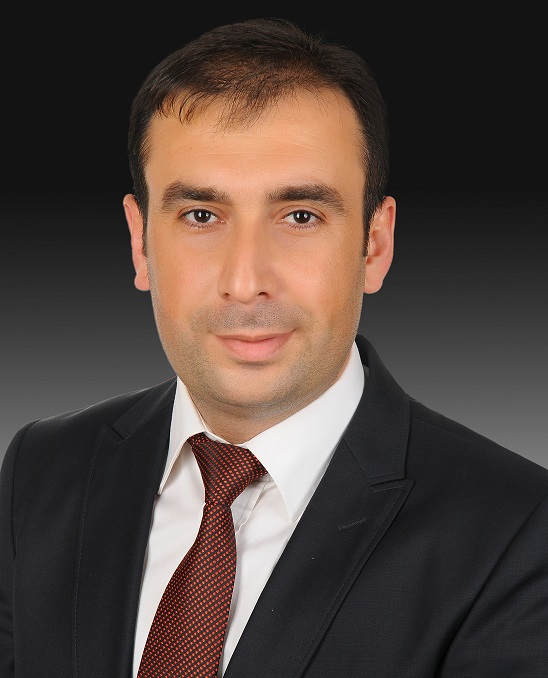}}]{Sabit Ekin (M'12)}  received the B.Sc. degree in electrical and electronics engineering from Eski\c sehir Osmangazi University, Turkey, in 2006, the M.Sc. degree in electrical engineering from New Mexico Tech, Socorro, NM, USA, in 2008, and the Ph.D. degree in electrical and computer engineering from Texas A\&M University, College Station, TX, USA, in 2012. He was a Visiting Research Assistant with the Electrical and Computer Engineering Program, Texas A\&M University at Qatar from 2008 to 2009. In summer 2012, he was with the Femtocell Interference Management Team in the Corporate Research and Development, New Jersey Research Center, Qualcomm Inc. He joined the School of Electrical and Computer Engineering, Oklahoma State University, Stillwater, OK, USA, as an Assistant Professor, in 2016. He has four years of industrial experience from Qualcomm Inc., as a Senior Modem Systems Engineer with the Department of Qualcomm Mobile Computing. At Qualcomm Inc., he has received numerous Qualstar awards for his achievements/contributions on cellular modem receiver design. His research interests include the design and performance analysis of wireless communications systems in both theoretical and practical point of views, interference modeling, management and optimization in 5G, mmWave, HetNets, cognitive radio systems and applications, satellite communications, visible light sensing, communications and applications, RF channel modeling, non-contact health monitoring, and Internet of Things applications.
\end{IEEEbiography}

\begin{IEEEbiography}[{\includegraphics[width=1in,height=1.25in,clip]{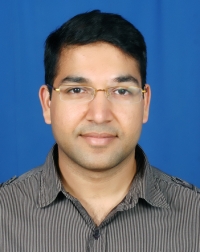}}]{Ali Imran (M'15)}  received the B.Sc. degree in electrical engineering from the University of Engineering and Technology, Lahore, Pakistan, in 2005 and the M.Sc. degree (with Distinction) in mobile and satellite communications and the Ph.D. degree from the University of Surrey, Guildford, U.K., in 2007 and 2011, respectively. He is an Assistant Professor in telecommunications with the University of Oklahoma, Tulsa, OK, USA, where he is the Founding Director of the Artificial Intelligence (AI) for Networks Laboratory (AI4Networks) Research Center and TurboRAN 5G Testbed. He has been leading several
multinational projects on Self Organizing Cellular Networks such as QSON, for which he has secured research grants of over 3 million in last four years as the Lead Principal Investigator. He is currently leading four NSF funded Projects on 5G amounting to over 2.2 million. He has authored over 60 peer-reviewed articles and presented a number of tutorials at international forums, such as the IEEE International Conference on Communications, the IEEE Wireless Communications and Networking Conference, the European Wireless
Conference, and the International Conference on Cognitive Radio Oriented Wireless Networks, on his topics of interest. His research interests include self-organizing networks, radio resource management, and big-data analytics. He is an Associate Fellow of the Higher Education Academy, U.K., and a member of the Advisory Board to the Special Technical Community on Big Data of the IEEE Computer Society.
\end{IEEEbiography}

\end{document}